**Data Descriptor:** A large-scale and PCR-referenced vocal audio dataset for COVID-19

Jobie Budd[1,2], Kieran Baker[3,4], Emma Karoune[4], Harry Coppock[4,5], Selina Patel[6,7], Ana Tendero Cañadas[6,8], Alexander Titcomb[6], Richard Payne[6], David Hurley[6], Sabrina Egglestone[6], Lorraine Butler[6], Jonathon Mellor[6], George Nicholson[9], Ivan Kiskin[10,11], Vasiliki Koutra[3,4], Radka Jersakova[4], Rachel A. McKendry[1,2], Peter Diggle[12], Sylvia Richardson[4], Björn W. Schuller[4,5,14], Steven Gilmour[3,4], Davide Pigoli[3,4], Stephen Roberts[4,9], Josef Packham[6], Tracey Thornley[6,15], Chris Holmes[4]

1. London Centre for Nanotechnology, University College London, UK
2. Division of Medicine, University College London, UK
3. King's College London, UK
4. The Alan Turing Institute, UK
5. Imperial College London, UK
6. UK Health Security Agency, UK
7. Institute of Health Informatics, University College London, UK
8. Centre for Stress and Age-Related Disease, School of Applied Sciences, University of Brighton, UK
9. University of Oxford, UK
10. University of Surrey, UK
11. The Surrey Institute for People-Centred AI, Surrey Sleep Research Centre, UK
12. University of Lancaster, UK
13. University of Cambridge, UK
14. University of Augsburg, Germany
15. University of Nottingham, UK

## Abstract

The UK COVID-19 Vocal Audio Dataset is designed for the training and evaluation of machine learning models that classify SARS-CoV-2 infection status or associated respiratory symptoms using vocal audio. The UK Health Security Agency recruited voluntary participants through the national Test and Trace programme and the REACT-1 survey in England from March 2021 to March 2022, during dominant transmission of the Alpha and Delta SARS-CoV-2 variants and some Omicron variant sublineages. Audio recordings of volitional coughs, exhalations, and speech were collected in the 'Speak up to help beat coronavirus' digital survey alongside demographic, self-reported symptom and respiratory condition data, and linked to SARS-CoV-2 test results. The UK COVID-19 Vocal Audio Dataset represents the largest collection of SARS-CoV-2 PCR-referenced audio recordings to date. PCR results were linked to 70,794 of 72,999 participants and 24,155 of 25,776 positive cases. Respiratory symptoms were reported by 45.62% of participants. This dataset has additional potential uses for bioacoustics research, with 11.30% participants reporting asthma, and 27.20% with linked influenza PCR test results.

| | |
|---|---|
| Measurement(s) | Vocal Audio (Cough, Speech, Exhalation); SARS-CoV-2 Infection Status; Influenza (A & B) Infection Status |
| Technology Type(s) | Microphone Device; PCR Test; LAMP Test, Lateral Flow antigen Test; Digital Survey |
| Factor Type(s) | Age; Gender; Ethnicity; Geographical Region; Symptoms; Respiratory Health Conditions; SARS-CoV-2 PCR Cycle Threshold Values; Influenza PCR Cycle Threshold Values; Smoker Status; Height; Weight; First Language |
| Sample Characteristic - Organism | Homo Sapiens |
| Sample Characteristic - Location | United Kingdom |

## 1. Background & Summary

The scale and impact of the COVID-19 pandemic has created a need for rapid and affordable point-of-care diagnostics and screening tools for infection monitoring. The possibility of accurate and generalisable detection of COVID-19 from voice and respiratory sounds using audio classification on a smart device has been hypothesised as a way to provide a non-invasive, affordable and scalable option for COVID-19 screening for both personal and public health monitoring[1]. However, prior machine learning studies to determine the feasibility of COVID-19 detection from audio have largely relied on datasets which are too small or unrepresentative to produce a generalisable model, or which include self-reported COVID-19 status, rather than gold standard PCR (Polymerase Chain Reaction) testing for SARS-CoV-2 infection (see **Table 1**). These datasets have a relatively small proportion of positive cases, and include inadequate metadata for statistical evaluation. They largely do not enable studies using them to meet diagnostic reporting criteria (for example the STARD 2015[2] and forthcoming STARD-AI[3] criteria), such as: reporting the interval between reference test and recording, random sampling, or avoiding case control where positives and negatives are sourced from different recruitment channels.

Following the publication of initial studies reporting accurate classification of SARS-CoV-2 infection from vocal and respiratory audio[4,5], the UK Health Security Agency (UKHSA, formerly NHS Test and Trace, the Joint Biosecurity Centre, and Public Health England) were commissioned to collect a dataset to allow for the independent evaluation of these studies. Dataset analysis was carried out by The Alan Turing Institute and Royal Statistical Society (Turing-RSS) Health Data Lab[6]. A dataset larger than the majority of existing datasets was needed to provide sufficient instances of various recording environments and mobile devices (information which is not collected), and to provide sufficient instances for the thousands of features or representations typically produced from short vocal audio samples[7]. Such a dataset also needed to be sufficiently large and diverse to validate model performance across various participant demographic groups and presentations of SARS-CoV-2 infection.

UKHSA developed an online survey to collect a novel SARS-CoV-2 bioacoustics dataset (**Figures 1.a, 1.b)** in England from 2021-03-01 to 2022-03-07[8] (**Figure 1.d)**, during dominant transmission of the Alpha and Delta SARS-CoV-2 variants and some Omicron variant sublineages[9]. Participants were recruited after undergoing testing for SARS-CoV-2 infection as part of the national "*Real-time Assessment of Community Transmission"* (REACT-1) surveillance study[10] and the NHS Test and Trace (T&T) symptomatic testing programme in the community (known as Pillar 2)[11]. To facilitate independent validation of existing models, audio samples common across existing studies were collected in the online survey, including: volitional (forced) cough, speech, and an exhalation sound. These were linked to SARS-CoV-2 testing data (method, results, date) for the test undertaken by the participant either as part of REACT-1 or T&T. Further data on participant demographics (age, gender, ethnicity, first language, location) and symptoms (type and date of onset) were collected in the online survey to monitor potential bias.

The UK COVID-19 Vocal Audio Dataset is designed for studies examining the possibility of classification of SARS-CoV-2 infection from vocal audio, including for the training and evaluation of machine learning models using PCR as a gold-standard reference test[12]. The inclusion of influenza status (for REACT-1 participants in REACT rounds 16-18) and symptom and respiratory condition metadata may provide additional uses for bioacoustics research.

**Table 1:** Summary of currently available COVID-19 biomedical acoustics datasets as of 2022-07-19

| **Dataset** | **COVID-Positive/ Total Participants** | **COVID Label** | **Modalities** | **Testing interval reported?** | **Metadata** |
|---|---|---|---|---|---|
| The UK COVID-19 Vocal Audio Dataset (this study) | 25,776/72,999 (35.31%) | PCR (93.25%) LAMP Lateral Flow | Cough, Exhalation, Speech | Yes | Symptoms; Respiratory Conditions; Age; Gender; First Language; Ethnicity; Smoker status; Ct values; Vaccination status; Influenza status; UK Region |
| Tos COVID-19[13] | 25,664/139,986 (18.33%) | PCR (19.36%) Lateral Flow | Cough | Yes (±3 days) | Symptoms; Close Contact; Risk group; Test location |
| COVID-19 Sounds[14] | 2,106*/36,116 (5.83%*) | Self-reported | Cough, Exhalation, Speech | Yes | Symptoms; Health conditions; Age; Sex; Smoker status; Language; Hospitalisation status |
| COUGHVID[15] | 1,010/20,072 (5.31%) | Self-reported, Clinician annotation | Cough | No | Symptoms; Respiratory conditions; Age; Gender; Location |
| Covid19-Cough[16] | 682/1,324 (51.51%) | PCR (28.85%), Self-Reported | Cough | No | Symptomatic |
| Coswara[17] | 389/2,233 (17.42%) | Self-reported | Cough, Exhalation, Speech | No | Health conditions; Symptoms; Age; Gender; English proficiency; Country; Locality; State; Smoker status; Vaccination Status |
| Virufy[18] | 143/456 (31.36%) | PCR (93.20%), Self reported | Cough | No | Symptoms; Medical conditions; Age; Sex; Smoker status |

*COVID-positive samples, individual COVID-positive participants may have recorded a sample more than once

## 2. Methods

### 2.1. Survey Design

Survey questions and responses are listed in **Supplementary Table 1**. Survey questions were designed to align to existing vocal acoustic data collection studies (see **Table 1**) and prevalence studies[19], so that future comparisons of study demographics could be made if necessary. These include variables that could be captured in vocal audio acoustic features and/or could confound with SARS-CoV-2 infection status[20,21], for example, respiratory symptoms, smoker status, and respiratory health conditions.

The participant's testing provider collected data on age, gender, ethnicity, geographical area and SARS-CoV-2 test result (and associated information such as test type and PCR cycle threshold information, if available). To minimise data entry fatigue, these were linked to survey responses and not collected again through the survey.

Survey variables were also chosen to align with existing government surveys for ease of comparison: options available for 'first language reflected those available in the ONS 2011 Census[22]; symptom options combined those available in the ONS Coronavirus Infection Survey[19] and the NHS Test and Trace symptom self-screening tool[23] (prior to April 2022). Additional symptom options were added 2021-07-21 ('other symptoms new to you in the last 2 weeks') and 2021-08-11 ('runny or blocked nose', 'sore throat') to capture symptoms reported at a higher frequency in COVID-19 variants circulating at the time[24]. All questions allowed a 'prefer not to say' option to maintain participant control on the data they chose to share and to minimise non-response bias.

The final survey questions requested participants to record short audio segments using the microphone on their device, where the user interface for making the recordings was embedded in the online survey. Audio recordings, in order of participant submission, were: a sentence read aloud, three successive *"ha"* exhalation sounds, one volitional cough, and three successive volitional coughs. Audio prompts were chosen to be similar to those of existing datasets (see **Table 1**), so that models trained on other datasets could be independently evaluated with this dataset. On completion of the survey, responses including audio data were sent to a secure server and temporarily held before being sent to UKHSA. Screenshots of the survey are shown in **Figure 1.b**.

Two cough recordings were captured of one and three successive coughs, matching the prompts for cough recordings captured in previous studies (see **Table 1**). A cough is an innate reflex to remove irritation in the respiratory tract, in order to enhance gas exchange. Coughs are typically associated with respiratory infection, and a new, persistent cough was one of three 'classic' COVID-19 symptoms, however it was less prominent compared to other respiratory symptoms in later variants[24]. The difference between a reflexive and volitional cough should be noted, where a volitional cough may differ in duration and power[25], and may be affected by the participants surroundings and emotional state. All coughs recorded in this study should be volitional, although a volitional cough may trigger a reflexive cough. Instructions were given to record the cough samples at an arm's length, following the advice provided to participants of the COUGHVID study[15], to reduce the risk of the audio recording being distorted (clipped). Participant instructions included guidance on coughing alone in a room or vehicle to reduce risk of COVID-19 transmission to others. Prompts and instructions are listed in **Supplementary Table 1**. The first (out of four total coughs) per participant may involve more fluid clearance. Of the successive (final three) coughs, the first was likely to be the most powerful, and successive coughs were likely to decrease in acoustic power as the participant had less time to inhale.

Exhalation sounds were also collected, as in previous studies. Breathing sounds are used in lung auscultation to identify narrowed airways or excessive fluid[26] in the respiratory tract, although the clinical utility of external recordings (without a stethoscope) has not been established. Participants are prompted to record three short, powerful exhalations (*"ha"* sounds, as if the participants "were trying to fog up a window, or see their[sic] breath in cold weather."). Participants were recommended to make this recording in a quiet environment to reduce background noise. For this recording, there was no direction around distance from participant to the recording device.

A sentence of speech, read from text, was also collected. Vowel sounds (such as 'aah' or 'ee') are used in lung auscultation (egophony) and to examine the vocal tract (such as contraction of the soft palate[27]). As speech is a combination of many varying vocal tract configurations over time, making it a more complex sample (anatomically) than coughing or breathing, it is more likely to be prone to biases in cognition, literacy, and accent and other learnt speech patterns. However, speech samples may potentially be more rich in acoustic features, particularly since smart device microphones, audio data processing, and the majority of vocal audio feature extraction models are configured for speech. Speech is produced through volitional manipulation of the vocal tract, where the shape of air cavities and air pressure is varied. A short sentence, *"I love nothing more than an afternoon cream tea"*, was chosen, combining several vowel and nasal sounds in a single recording.

## 2.2. Recruitment

Participants were recruited through two existing SARS-CoV-2 infection testing pathways in parallel: 1) a community prevalence survey and 2) a government testing service. They were invited to take part in the study after they underwent testing. Survey responses and audio recordings were then linked to their test result. Inclusion criteria across both recruitment channels were: being 18 years of age or older and having a COVID-19 test barcode number. Participants were also advised to participate only if they had tested in the last 72 hours, although 13.13% of REACT-1-recruited and 2.12% of NHS Test and Trace-recruited participants in the dataset have a submission delay exceeding 72 hours, see **Figure 1.e.** (submission delay is described in the participant metadata file, see **Supplementary Table 3)**. Participation was completely voluntary. This study includes some participants with a permanent address in the UK devolved administrations (Northern Ireland, Scotland, Wales), however, the data is disproportionately England-sampled due to the England-only recruitment of the majority of recruitment routes.

Participants were recruited via the Real-time Assessment of Community Transmission-1 (REACT-1) study[10]. REACT-1 was commissioned by the UK Department of Health and Social Care to estimate the prevalence of SARS-CoV-2 infection in the community in England (and influenza A and B in later survey rounds). It was carried out by Imperial College London in partnership with Ipsos MORI using repeat, random, cross-sectional sampling of the population. Participants were randomly selected from National Health Service England records (which include almost the entire population) and sent a letter invitation, with the aim of creating a representative sample of the population for each survey round (although actual response demographics vary, see **Section 5**). Participants were provided with instructions to take a throat and nasal self-swab and were asked to respond to an online/telephone survey about their demographics, symptoms and recent behaviours. The swab was either posted or collected by courier for PCR testing at laboratories. For rounds 13-18 (REACT survey dates from 2021-06-24 to 2022-03-01), participants were asked if they agreed to be contacted about further research led by the UKHSA. After sending their swab to a laboratory and completing the REACT-1 survey, those who agreed to be contacted were sent an email invitation to the online survey for this study which included audio recordings (survey questions and responses listed in **Table 2**). 12.22% of the 295,493 individuals contacted for recruitment in REACT rounds 14-18 participated in the study and are included in the final dataset. **Supplementary Table 2** lists participant cohorts and recruitment methods in further detail.

Participants were also recruited via SARS-CoV-2 testing services delivered by NHS Test and Trace (T&T). The purpose of this recruitment channel was to increase the number of survey responses linked to a positive PCR test result to better balance the combined dataset by SARS-CoV-2 infection status. Where the prevalence of SARS-CoV-2 infection of the REACT-1 cohort was expected to be similar to the prevalence in the general population, a higher proportion of positive cases may be needed for the development of SARS-CoV-2 infection status classification models. During the study period, people were advised to seek a PCR-test through T&T (swab testing for the wider population, as set out in government guidance, known as Pillar 2[11]) if they were (i) experiencing COVID-19 symptoms, (ii) identified as a close contact of a positive case, or (iii) taking a confirmatory PCR test following a positive rapid antigen (lateral flow) test (until 11th January 2022). Tests were free to use and available at test sites or for home delivery. Throat and nasal swabs were mostly self-administered at test sites or in participants' homes, before being sent to laboratory sites for testing[11].

A subset of participants recruited through T&T reported lateral flow test results. Lateral flow testing of SARS-CoV-2 antigen was open and free to the public, including for asymptomatic testing, and in the majority of cases was performed by the participant and reported through the NHS COVID-19 app or website.

Those that underwent testing could opt-in to be contacted about participating in research. An eligible subset of these were then contacted by text, email or phone call to invite them to participate in the study. **Supplementary Table 2** lists participant cohorts and recruitment methods in further detail. Eligible populations were defined by SARS-CoV-2 infection and symptom status over a distribution of ages. Recruitment was initially focused on those receiving a positive test result. Between 2021-11-11 and 2022-03-04 recruitment was targeted at 50% of a random sample of all that includes those testing positive, negative or with a void PCR test result. Participants were linked to an online survey where prompts and audio recording format were uniform across recruitment channels.

A small proportion of participants prior to 2021-03-17 were recruited via information leaflets at regional COVID-19 test sites, displaying a QR code linked to the study survey.

Participants were also recruited from The ONS Coronavirus Infection Survey[28] and the COVID-19 Challenge study (COV-CHIM 01)[29], however lower participant counts from these recruitment methods could not guarantee participant anonymisation and so these participants are not included in the UK COVID-19 Vocal Audio Dataset.

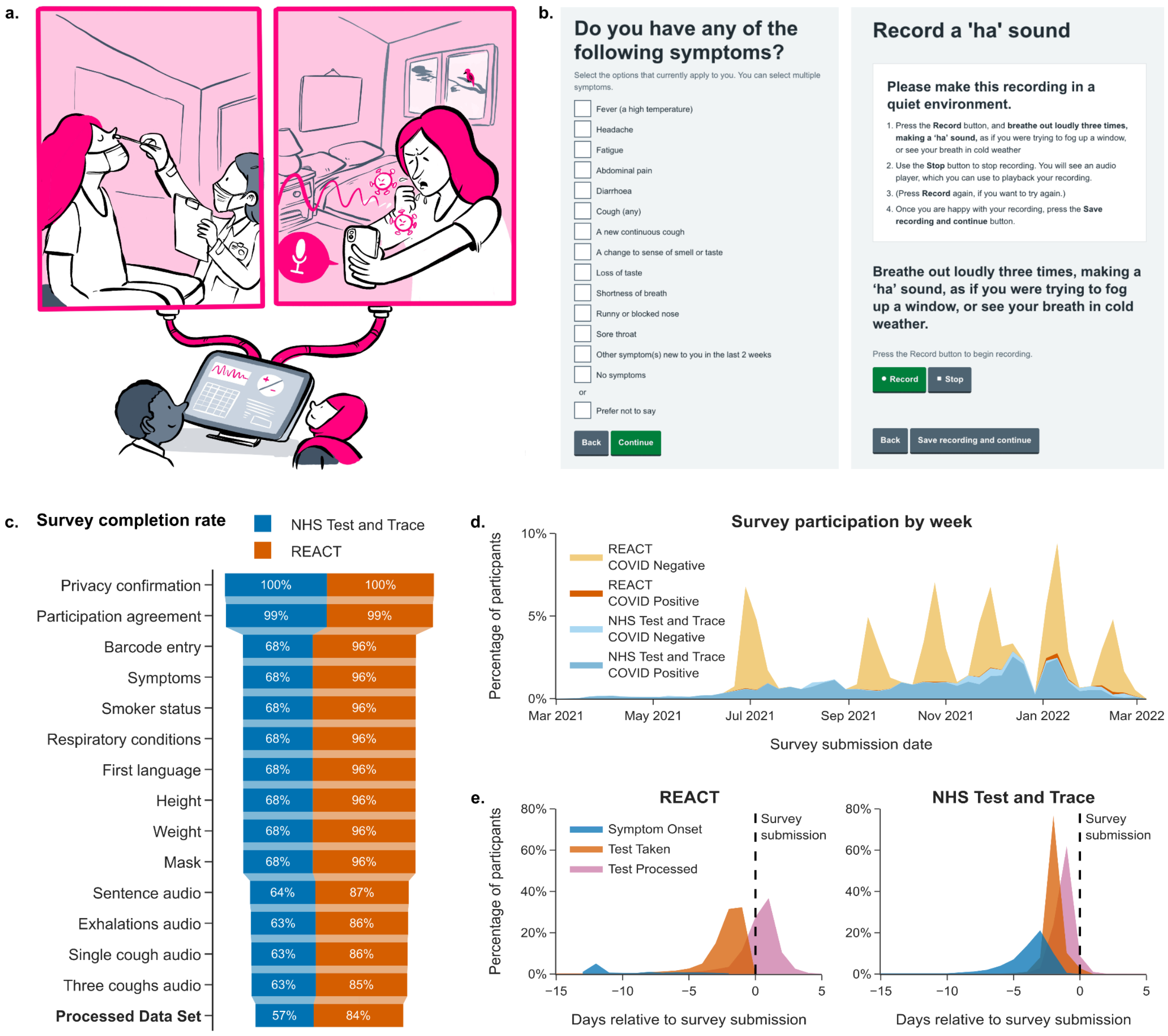

***Figure 1. Study recruitment: a.** Illustration of dataset components. Participants of the REACT-1 study, or NHS Test and Trace patients underwent a test for SARS-CoV-2 infection. In this illustration the swab is healthcare worker-administered, however the majority of swab samples were self-administered in this study. Individuals from both cohorts were contacted and prompted to complete a digital survey, including recording a volitional cough and other respiratory sounds. SARS-CoV-2 test result data and associated information was combined with survey results and audio recordings, and was de-identified. This data descriptor document describes the process of producing the combined dataset, and its contents. This illustration is created by Scriberia with The Turing Way community (used under a CC-BY 4.0 licence. DOI: 10.5281/zenodo.3332807). **b.** Screenshots of the 'Speak up to help beat COVID' digital survey. **c.** Participant survey completion rate by survey question across both recruitment modes. Completion counts were only collected for the 'beta' survey phase, described in Section 2.3 (with 59,431 participants equalling 81.41% of total dataset). **d.** Participant records as a percentage of dataset total by week of survey submission, recruitment source, and SARS-CoV-2 test result. Individual REACT-1 survey rounds can be seen as peaks at irregular intervals. **e.** Time of symptom onset, SARS-CoV-2 test swabbing (test start date), and SARS-CoV-2 test processing in relation to time of study survey submission in days, for each recruitment source. Percentages shown as the dataset total for each recruitment source. Symptom onset records shown only where symptoms were*

*reported. Participants had been made aware of their SARS-CoV-2 test results on or shortly after the test processed date. REACT-1 participants who had an influenza test would have completed the test swab on the same date as their SARS-CoV-2 test, but would not have been made aware of the result[30].*

## 2.3. Data Collection

The online survey 'Speak up and help beat coronavirus'[8] was accessible via compatible internet connected devices with the ability to capture audio recordings, such as smartphones, tablets, laptops, and desktop computers. Participants recruited from T&T testing services were contacted to take part in the study after completing a SARS-CoV-2 test and agreeing to take part in research (see **Supplementary Table 2** for modes of contact). Those recruited from the REACT-1 cohort were sent an email invitation. Participants reviewed the participant information and confirmed their informed consent to take part. An automated check to confirm a participant's device was able to record audio was integrated into the digital survey, which participants needed to complete before continuing. Participants accepted a privacy statement outlining how their survey and test data would be linked, how their data would be used for research, and made available for reuse by researchers. Next, they entered their test/personal barcode number, followed by responses to questions about their demographics, comorbidities and any symptoms they were currently experiencing. Participants responded to survey questions from a choice of predefined responses (survey questions and multiple-choice responses listed in **Supplementary Table 1,** survey completion rates listed in **Figure 1.c**).

Until 2021-08-12, the 'alpha phase' gathering solution was hosted at www.ciab2021.uk (used by 18.59% of participants, noted as '`alpha`' in the '`survey_phase`' metadata variable, see **Supplementary Table 3**), and from 2021-08-13 to 2022-03-07, the 'beta phase' data gathering solution was hosted at www.speakuptobeatcovid.uk (used by 81.41% of participants, noted as '`beta`' in the '`survey_phase`' metadata variable, see **Supplementary Table 3**). To ensure robustness, both data gathering solutions were tested extensively to ensure data gathered was recorded accurately in the databases of the respective solution, and to confirm that the data was subsequently transferred to UKHSA correctly. This included end-to-end tests including dummy submissions.

The API and associated configuration used for recording audio in the 'alpha' solution was replicated as-is in the 'beta' solution. Recordings through both data gathering solutions were compared to check consistency, including comparison of Fast Fourier Transform (FFT) spectra, file format, and sampling rates (using the python librosa library[31]). The solution delivery teams for both 'alpha' and 'beta' data gathering solutions confirmed that no post-processing of the stored audio files occurred for either solution.

## 2.4. Data Linking

A data pipeline was designed to merge the primary data gathered in support of this study (the submission data) with the secondary data (the SARS-CoV-2 test results data) gathered by each testing provider. Data pipeline code was drafted and peer-reviewed by the UKHSA study team, and was also reviewed independently by the data wrangling team from Turing-RSS. Survey data and audio recordings submitted by the participant were linked to the SARS-CoV-2 test result data (date, result, test type, testing laboratory, PCR cycle threshold values (if provided), and estimated viral load (if provided)) for the test they underwent prior to being recruited. They were also linked to demographic information of potential additional utility to the dataset (age, gender, ethnicity, geographical information, COVID-19 vaccination status) which was collected by the testing provider.

Test barcodes were used to link T&T data to survey data. Test results data from T&T-recruited participants were sourced from the National Pathology Exchange (NPEx) database that stored test result data from across the T&T laboratory network and home-based lateral flow test results. Participant age, gender, ethnicity, and location were derived from data entered by the participant when registering for a test and stored in the NPEx database.

The study team generated a set of unique personal codes, which were provided securely to Ipsos MORI, who included a code in each email invitation to participate in this study. These personal codes differed in format from T&T barcodes to avoid accidental duplication. This personal code was used to link REACT-1 data to survey data. For REACT-1-recruited participants, the test result and associated data were provided by Ipsos MORI as a filtered extract of the REACT-1 study data including only the records and fields relevant to this study. Participant codes were extracted from responses to the survey for this study and transferred via approved protocols to Ipsos MORI. Ipsos MORI checked for duplicate entries before then extracting and sharing the test result data for UKHSA to link back to survey submissions.

The pipeline script was designed to exclude any participant submissions from the final dataset that could not be linked to valid test results data using the test identifier code submitted by the participant. The test identifier codes were not publicly available and were provided to the participant through the relevant recruitment route, mitigating the risk of any submissions where the primary data gathered was provided by a different individual from the secondary test results data.

To further mitigate this risk, as well as provide a metric required for the study exclusion criteria, the pipeline script calculates the delay between the time of the participant's submission to the primary data gathering solution, and either the swab time or lab processing time for the SARS-CoV-2 test the participant conducted ('`submission_delay`' variable, see **Supplementary Table 3**). This delay was calculated as the difference between the time of the participant's submission to the primary data gathering solution, and either the swab time or lab processing time for the test ('`submission_delay`' variable, see **Supplementary Table 3**). This variable enables results to be filtered out from the study data if there was a significant delay between submission and SARS-CoV-2 test, as this could either suggest the participant entered the test identifier incorrectly and has been associated with the wrong test result record, or due to the delay the test may no longer be indicative of the participant's SARS-CoV-2 infection status.

Participant submissions which could not be linked to a valid SARS-CoV-2 test result were excluded from the final dataset. Test barcode numbers were removed at the end of the study, and replaced with a random identifier associated with an individual participant (‘`participant_identifier`’ metadata variable, see **Supplementary Tables 3, 4**), de-linking the participant metadata from their test barcode number and identifiable information associated with it.

## 2.5. Data Cleaning

Duplicate entries from the same participant were removed where possible so that the data tables have one row (equivalent to one survey entry) per participant. For T&T-recruited participants, repeat individuals were identified in the source test results data table, and repeat submissions were removed keeping only each individual’s first submission, which would be closer to the time of the participant’s SARS-CoV-2 test. For the REACT-1-recruited participants, Ipsos MORI indicated which submissions related to individuals who had previously taken part in the study, and repeat submissions were removed keeping only first submissions. There remains a residual risk that some individuals took part in both recruitment groups and as a result have made multiple submissions in the study data, however, this is expected to be a low volume due to the national scale of both recruitment approaches. The removal of duplicates cannot be guaranteed prior to 2021-06-01 (1.38% of participants), as a shorter agreed personal data retention period for this ‘pilot’ phase of data collection meant that test barcodes could not be stored for the duration of the study.

Variable categories were standardised for uniformity across recruitment channels where there was overlap between categories. REACT category names were typically renamed to match T&T category names. Data types were standardised by variable (unless disclosure controls required mixed data types). List variables from survey multiple choice questions were one-hot encoded. Geographical information was originally collected at the local authority (sub-regional administrative division) level, and was later aggregated to region (first level of national sub-division) level to avoid the risk of participant disclosure. A `pseudonymised_local_authority` variable remains which can be used to stratify participants by geography for model evaluation. First language values with total counts <5 were pseudonymised. All remaining variable category values were grouped to remove low counts (<5) to reduce the risk of participant disclosure.

## 2.6. Data Anonymisation

To enable wider accessibility, an open version of the COVID-19 Vocal Audio Dataset was produced according to the ISB1523: Anonymisation Standard for Publishing Health and Social Care Data standards[32]. To meet these standards, the audio recordings of read sentences were removed, and several participant metadata variables were either removed, binned, obfuscated, or pseudonymised to meet the requirement of K-3 anonymity after combining all variables relating to personal data. A data dictionary for the reduced participant metadata is provided in **Supplementary Table 5.**

## 3. Data Records

The open version of the COVID-19 Vocal Audio Dataset is available for download[33]. Additional data records are available for access via application to UKHSA (see **Section 5**). There were 72,999 participants included in the final dataset, with one submission per participant. This included 25,776 participants linked to a positive SARS-CoV-2 test. The majority of these submissions (70,794, 96.98%) were linked to results derived from PCR tests (RT-PCR, q-PCR, ePCR) , followed by lateral flow tests (1,925, 2.64%) and LAMP tests (244, 0.33%). Of all test results, 178 (0.24%) were inconclusive with an unknown or void result. This dataset represents the largest collection of PCR-referenced audio recordings for SARS-CoV-2 infection to date, with approximately 2.6 times more participants with PCR-referenced audio recordings than the Tos COVID-19 dataset (with 27,101)[13].

The majority (44,565 participants, equalling 61.05% of total dataset) of participants were recruited via REACT-1. The remaining 28,434 participants, (equalling 38.95% of the total dataset) were recruited via T&T testing services. **Figure 2** shows a summary of participant attributes. The median age of participants was 53 years old and 59.64% of participants were female (43,537 participants). Participants with a positive SARS-CoV-2 test result were more likely to report that they were experiencing respiratory symptoms (90.73% of participants testing positive reported respiratory symptoms vs. 20.89% of participants testing negative reported respiratory symptoms). This dataset has additional potential uses for bioacoustics research, as 9,749 (13.35%) participants reported a pre-existing respiratory health condition (comorbidity) of which 8,249 (11.30%) reported asthma. Participants recruited from REACT rounds 16-18 (19,859 participants, 27.20% total) have linked influenza PCR test results, with 33 participants testing positive for influenza A and 28 testing positive for influenza B. Recruitment was focused in England, however several participants had a permanent address in Wales or Scotland. Several recruitment biases were apparent and are discussed in **Section 5**.

Audio was recorded in *.wav* format (86.12% of submissions had a sample rate for all recordings of 48kHz, 13.14% of submissions had a sample rate for all recordings of 44.1kHz) and had a maximum length of 72 seconds (see **Figure 3.A**). Four audio files (one for each recording) are provided for each of the 72,999 participants (unless missing, see **Section 4**).

Metadata including audio filenames are provided in three .csv files, linked by a participant identifier code. Metadata data dictionaries are provided as tables for participant metadata (**Supplementary Table 3**) and audio metadata (**Supplementary Table 4**).

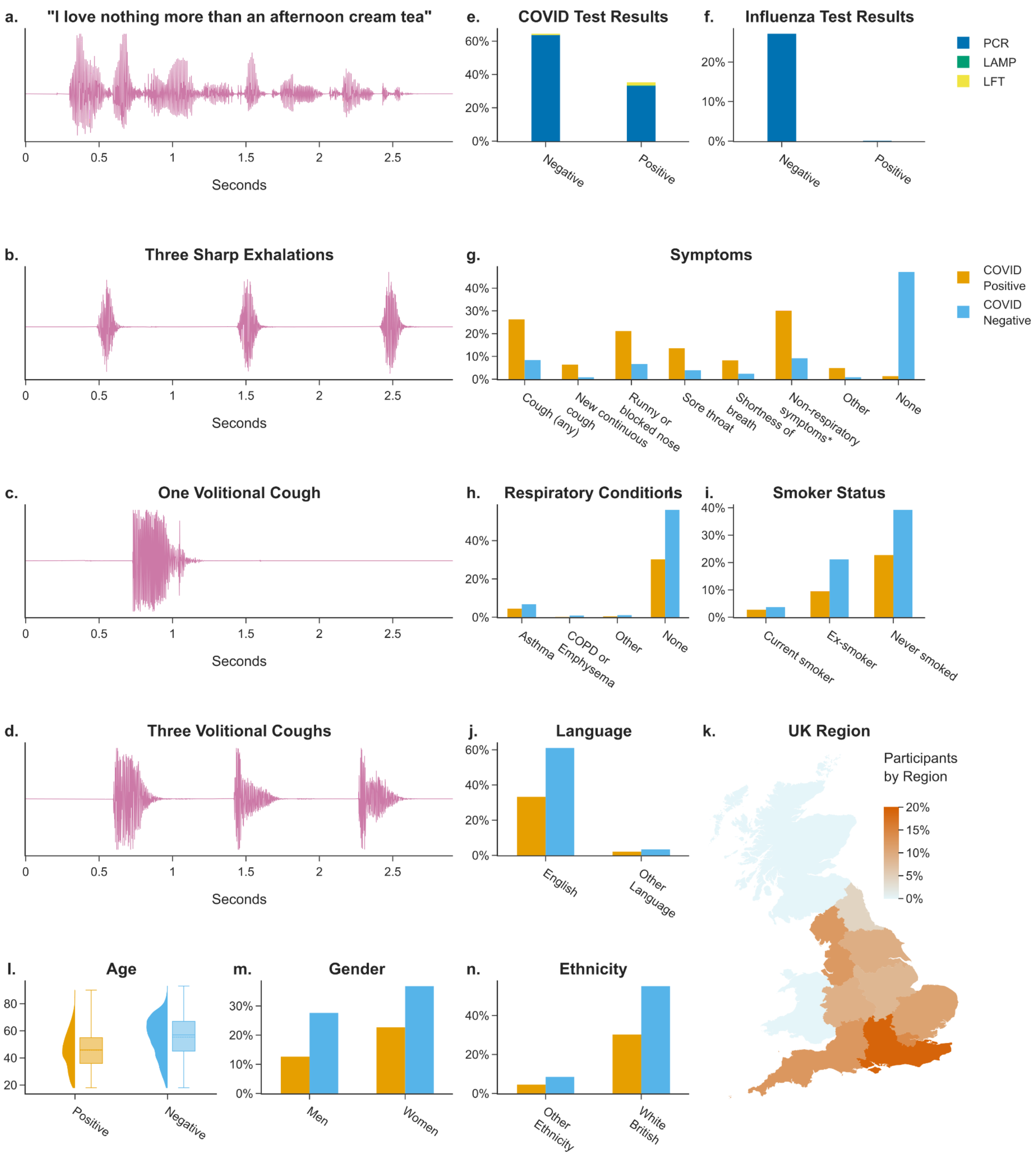

***Figure 2**. **Dataset summary:*** *All percentages from the dataset total (72,999 participants). 'Prefer not to say' or missing responses are not visualised (variable completeness statistics are listed in Supplementary Tables 3 and 4). Multiple categories with low counts grouped for visualisation.* ***a-d*** *Waveforms of the four audio recordings captured for each participant (a read sentence, three successive exhalations, one volitional cough, three successive volitional coughs; here recorded as an example by the primary author at 44.1 kHz when asymptomatic and with unknown SARS-CoV-2 infection status) .* ***e.*** *Percentage of total participants by COVID-19 status and test type.* ***f.*** *Percentage of total participants by Influenza status (A or B) and test type.* ***g.*** *Percentage of total participants by self-reported symptoms and SARS-CoV-2 test result. *Non-respiratory symptoms include fever or high temperature, headache, abdominal pain, loss of taste, and changes to sense of smell or taste.* ***h.*** *Percentage of total participants by self-reported respiratory condition and SARS-CoV-2 test result.* ***i.*** *Percentage of total participants by smoker status and*

*SARS-CoV-2 test result.* ***j.*** *Percentage of total participants by first language and SARS-CoV-2 test result.* ***k.*** *Geographical distribution of participants by UK administrative region (2021 boundaries*[34]*, no participants were recorded as resident in Northern Ireland, Shetland (Scotland) not shown)* ***l.*** *Participant age distribution by SARS-CoV-2 test result.* ***m.*** *Percentage of total participants by gender (REACT-1/T&T categories) and SARS-CoV-2 PCR test result* ***n.*** *Percentage of total participants by ethnicity (REACT-1/T&T categories) and SARS-CoV-2 PCR test result.*

## 4. Technical Validation

Audio *.wav* files were parsed and metadata extracted including sample rate, number of samples, and number of channels (using the python scipy library[35]). Using the audio data and extracted metadata, the audio length in seconds, audio amplitude (absolute maximum - absolute minimum signal), and audio signal-to-noise ratio were calculated for each file. **Figures 3.a-c** show the distribution of audio file length, audio amplitude, and audio signal-to-noise ratio for each audio recording type, respectively. Here, signal-to-noise ratio is the ratio between the absolute signal mean and the absolute signal standard deviation, and does not distinguish vocal signal from background noise. All files had one audio channel. Audio metadata is available in the `audio_metadata_df` table of the dataset. 2.47% participants were missing one or more audio files, or had audio files with a size of <45 bytes, and were flagged with the `missing_audio` variable.

Audio files were screened systematically to reduce the risk of disclosure of personal information. This could arise where participants had failed to follow the study instructions and the audio prompts, instead accidentally or intentionally disclosing personal information such as their name. An analytical pipeline was developed to identify outliers from the total 289,696 audio files, where the outliers were screened manually. A speech-to-text model (fairseq S2T, small version, pre-trained weights available at https://huggingface.co/facebook/s2t-small-librispeech-asr)[36] was run on audio files, producing a text transcript. A text-to-embedding model (MPNet Transformer v2, pre-trained weights available at https://huggingface.co/sentence-transformers/all-mpnet-base-v2)[37] was run on the transcript, producing a format of the transcript (an embedding) that was quantified and compared with the prompt to identify outliers, or speech that differs from the prompt.

Each embedding was then ranked by its similarity to the prompt using a Support Vector Machine (SVM) model. The 1000 sentences which differ most were then inspected manually to check for disclosure of personal information, and non-outlier files were randomly sampled. **Figure 3.d** shows the distribution of the similarity rank for every audio file for the sentence modality. The majority of transcripts (56.05%) show the correct sentence (*"I love nothing more than an afternoon cream tea"*). Most others sampled show a similar sentence, misinterpreted by the speech-to-text model, with 95.63% of transcripts containing the substring *"nothing more"*, and 91.84% containing *"nothing more"* and *"afternoon"*, with the other words commonly mis-transcribed. A small proportion of sampled outlier transcripts show alternative speech from the participant. These transcripts and associated audio files were retained unless personal information was disclosed (one audio file was found to contain personal information and was truncated). Other outliers had media playing in the background, or others show artefacts of noise e.g. "*of of of of of of of of of…*". Sentence transcripts and outlier scores are available in the `audio_metadata_df` table of the dataset.

Several data filtering steps are recommended when using this dataset for the development of models with the intention of SARS-CoV-2 infection classification, including filtering to include only participants with a PCR-type SARS-CoV-2 test, participants whose test was not carried out in a laboratory with reported testing errors, and participants who completed the study survey within a defined delay (e.g. 10 days) of their SARS-CoV-2 test. *Pigoli et al.* describe these suggested data filtering steps in further detail[38].

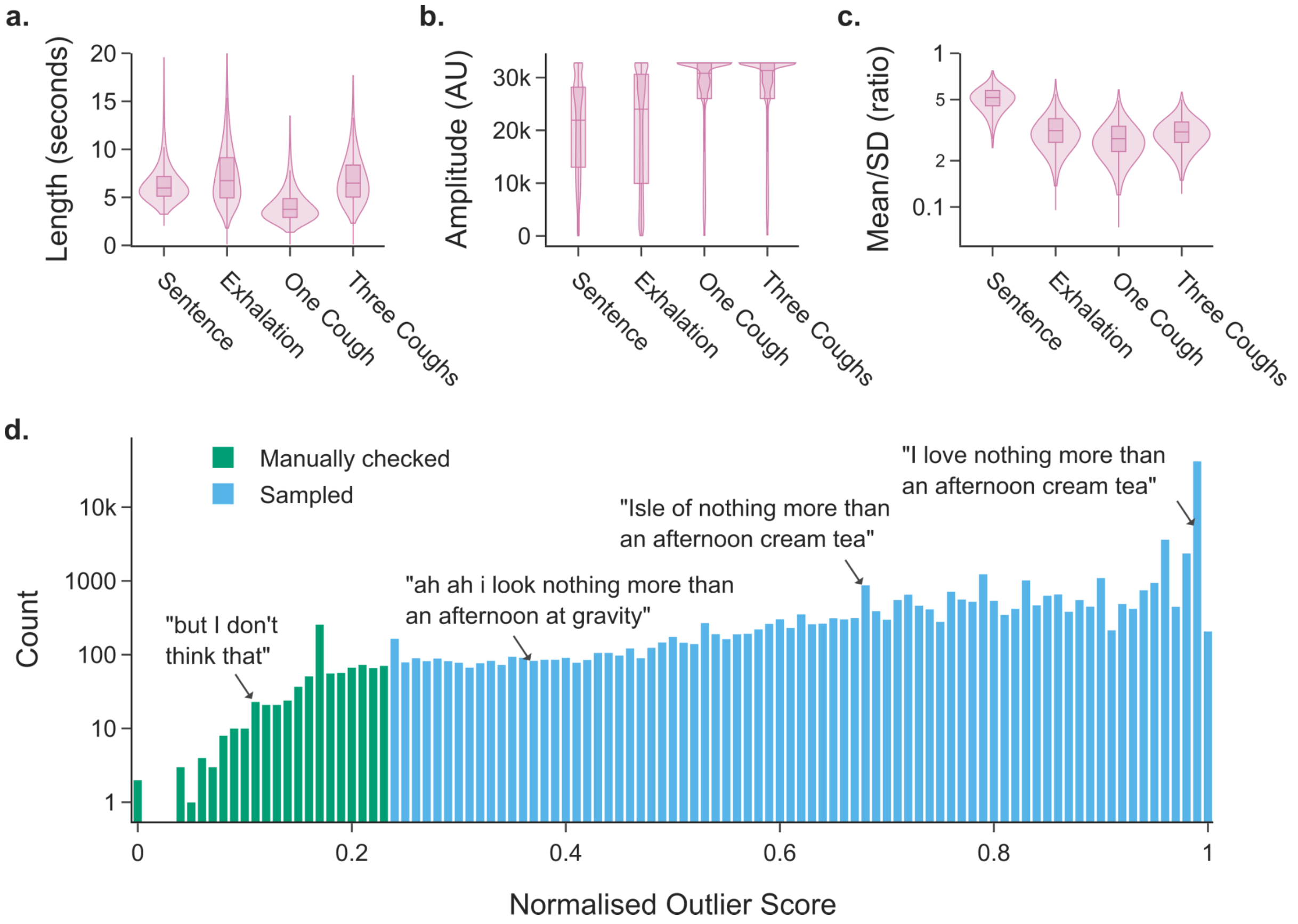

***Figure 3. Audio data technical validation:*** ***a.*** *The distribution of audio clip length in seconds for each audio modality.* ***b.*** *The distribution of audio clip amplitude (difference between maximum and minimum of absolute signal) for each audio modality.* ***c.*** *The distribution of audio clip absolute signal means divided by absolute signal standard deviation for each audio modality;* ***d.*** *Normalised anomaly scores of sentence audio clip transcript embeddings, where the 1000 most anomalous were checked manually for personal information.*

## 5. Usage Notes

There is some bias in the recruitment for this study, where the majority of participants recruited via the REACT-1 surveillance study were SARS-CoV-2 negative at the time of participation, and the majority of the participants recruited via T&T were SARS-CoV-2 positive at the time of participation. This selection was necessary to produce a dataset with a relative balance of SARS-CoV-2 infection status, but researchers should note the varying composition of each recruitment population, which could confound with infection status. Additionally, within T&T recruitment, the recruitment method of some positive and negative cases varied (see **Supplementary Table 2**). *Pigoli et al.* document the potential confounders[38]. Of particular note is symptom presentation, where participants recruited via T&T would have sought a PCR test due to having a positive lateral flow test (between 2021-03-29[39] and 2022-01-11[40]) or having symptoms (at least one of: a high temperature; a new continuous cough; a change to sense of smell or taste) as per UKHSA guidance at the time of data collection[11]. Changes in UK testing policy, such as local surge testing or school and workplace testing policies would also create selection biases for the T&T population that varied over time. Compared to T&T, COVID-19 positive participants recruited via the REACT-1 study were less likely to be symptomatic. The distribution of symptom status was more likely to reflect that in the general population and be stable over time (in relation to SARS-CoV-2 prevalence). Users should note that not all those who were contacted to participate in the REACT-1 survey participated, creating some self-selection biases. Only those who participated were contacted to participate voluntarily in this study, leading to further self-selection biases.

Demographic imbalances are present within the dataset, where study participants were more likely to be White British, women, and aged 35-74 years than the general UK population. **Figure 4** compares the distribution of ages, genders, ethnicities, and region of habitation of study participants in comparison to the general population (as recorded by the 2021 UK Census), patients using T&T in the weeks of data collection[41] (compared to only T&T-recruited study participants), and REACT-1 study participants for the relevant study rounds[42–45] (compared to only REACT-1-recruited study participants). Some dataset biases can be seen to be partly inherited from the two recruitment channels, as in the case of gender, where more patients or participants were women in comparison to the general population. Other biases, such as age biases, can be seen to be exacerbated by the recruitment of this study, where fewer participants over the age of 80 years were seen in the two recruitment channels (and none under the age of 18 years, due to study exclusion criteria). The survey for this study was only made available in English which could have exacerbated language and ethnicity biases. Similar demographic biases were present in other voluntary digital surveys for COVID-19 research and surveillance in the UK, such as the COVID Symptom Study[46]. The nature of study recruitment and participation may exclude certain demographic groups with limited digital literacy or access to digital infrastructure[47]. The voluntary nature of this study may exclude certain demographic groups with limited available time due to employment and/or care commitments[48]. We recommend that researchers using this dataset to train audio classification models should report test accuracy statistics stratified by demographic variables to communicate any model biases.

The participant metadata variables not captured directly in the digital survey (digital survey questions and related variables listed in **Supplementary Table 1**) were shared by the relevant recruitment channel (see **Section 2.1**), where format and prompt vary. Efforts have been made to standardise data format between recruitment channels and are listed in the participant metadata dictionary (**Supplementary Table 3**). Users should note that some calculated variables, such as `symptom_onset`, continue to have values of distinct distributions despite this standardisation due to the variation in recruitment methods (patients seeking a test vs survey population). T&T- and REACT-derived demographic variables had limited multiple-choice options and limited ethnicity and gender categories were available, meaning some demographic analyses are not possible.

The substantial majority of participants (94.51%) report English as their first language or the language most commonly spoken at home (if they have two or more first languages). Therefore, any analysis of speech data may only be valid in English speakers and should be tested in other populations before language-generalisable results are reported. Regional accents may have an effect on speech models. Recruitment is relatively balanced by administrative region, particularly for REACT-1-recruited participants. As a result, the audio data may contain a representative sample of regional English accents. Most study participants were recruited in England and so, more data would be needed to evaluate accents which are more common outside of England.

Participant height and weight variables were selected from drop-down selections of value ranges and units. A disproportionate percentage (1.08% height of <=90 cm, 0.79% weight of 20kg) of participants submitted the lowest value, which was the default selected value before an option was made. It is likely that a proportion of participants selecting these lowest values have not submitted their true heights or weights.

*Coppock et al.* list seven core issues with existing COVID-19 audio research and the datasets used[49]. While we have designed this dataset attempting to address these issues; including providing PCR-confirmed infection status, providing demographic and health metadata for each participant, ensuring only one submission per participant, and publishing this dataset; several issues remain. Participants testing positive for SARS-CoV-2 infection at the time of participation may be aware of their infection status, particularly those recruited via T&T (**Figure 1.e**). This may introduce undocumented confounders in the audio recordings, such as behaviour when recording, the environment in which recordings are made, and participant emotions. Recruitment biases, discussed above, may also present undocumented confounders present in the audio data. Although there is little variation in audio sample rate across the dataset (see **Section 3**), we did not record device type, microphone hardware specifications, browser, or device operating system, which may have some effect on audio quality.

Although PCR is the gold-standard for detection of SARS-CoV-2 infection, users should note that it may be an imperfect label in categorising participants as infectious. Due to the amplification step, viral RNA can remain detectable by PCR long after live SARS-CoV-2 can be cultured from patient samples. Stratifying model evaluation by estimated viral load may remedy this, where studies have shown that the viral load threshold for transmission is ~1,000,000 viral RNA copies/ml[50]. We include viral load data where available (14.62% of submissions). Users are encouraged to use the `covid_viral_load_category` variable rather than the `covid_viral_load`, `covid_ct_value` or `covid_ct_mean` variables, as there is variation between tests including different gene targets (documented by

covid_ct_gene). False negative PCR results are also possible, likely related to sampling technique, volume of fluid, and viral load. A meta-analysis found a pooled estimate of 94% PCR sensitivity[51]. PCR results from a laboratory reported to have made substantial testing errors are flagged in the covid_test_lab_code variable.

The period of data collection saw different SARS-CoV-2 variants (notably Alpha, Delta and Omicron) circulating in the UK, which have been reported to cause differing prevalence of symptoms to each other and to previous variants[24]. Dataset authors recommend against using the covid_ct_gene metadata variable to estimate SARS-CoV-2 variant causing infection (e.g. through S-gene dropout[52]), as this variable reports only a single gene target with the lowest cycle threshold value, and not all laboratories test for all genes.

Due to recruitment constraints, we were unable to include longitudinal data (multiple data entries by the same participant over time) for any participants, as is present in other datasets such as the COVID-19 Sounds[14] dataset. As a result, this dataset is insufficient to study potential vocal changes throughout SARS-CoV-2 infection in the same individual. Temporal changes may be studied in a cross-sectional manner with appropriate controls using the symptom_onset variable.

## Data Access

The open version of the UK COVID-19 Vocal Audio Dataset[33] (see **Section 2.6**) is available under an Open Government License (v3.0)[53]:
https://zenodo.org/doi/10.5281/zenodo.10043977

Access to the full dataset may be requested from UKHSA (DataAccess@ukhsa.gov.uk), and will be granted subject to approval and a data sharing contract. To learn about how to apply for UKHSA data, visit:
https://www.gov.uk/government/publications/accessing-ukhsa-protected-data/accessing-ukhsa-protected-data

Additional data available include participant train/test splits for the experiments reported by *Coppock et al.*[12] (in both versions of the dataset) and OpenSmile features[7] generated from the audio files (in the full dataset only).

Access to a GPU is recommended for training machine learning algorithms with audio data from this dataset.

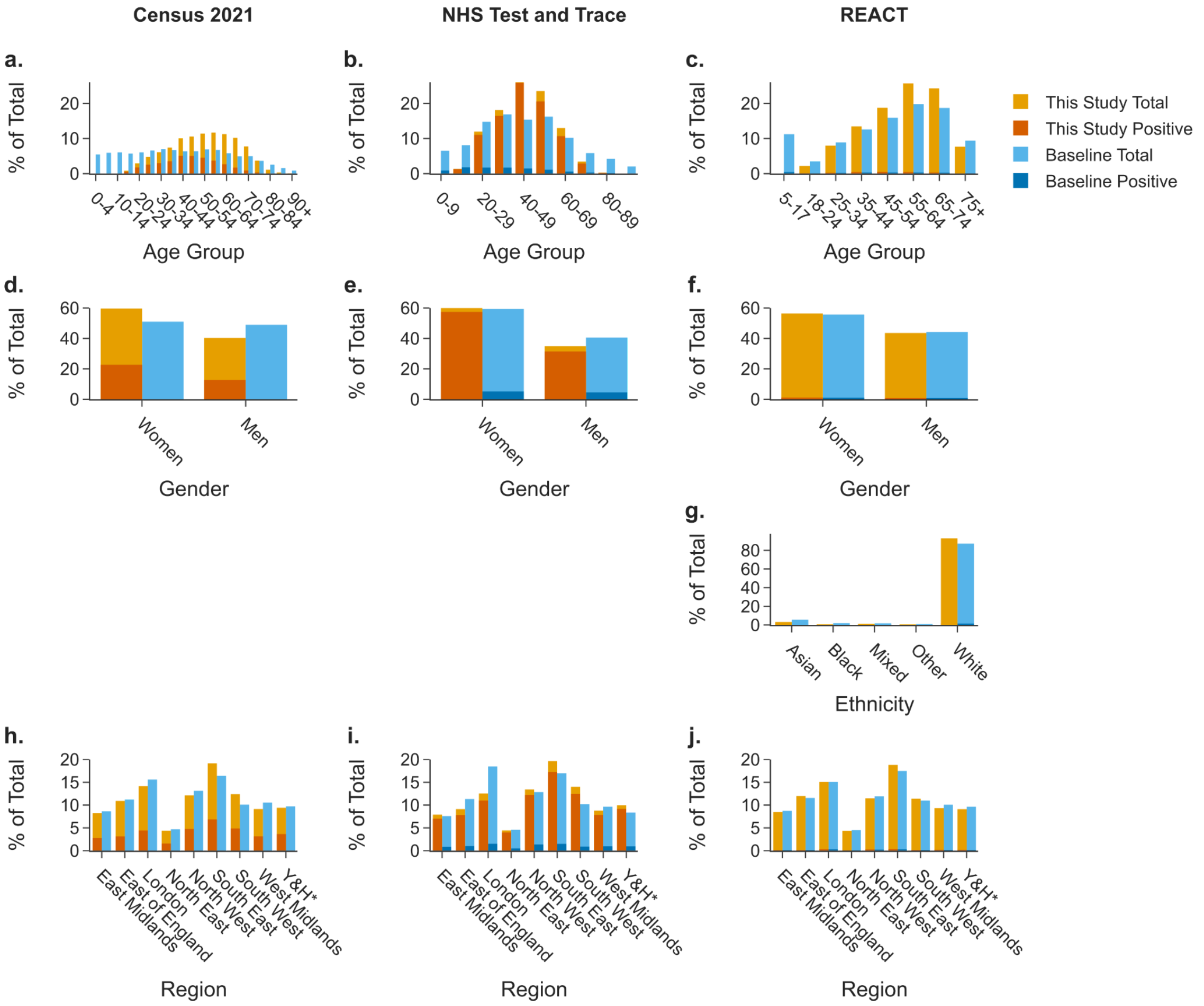

***Figure 4: Demographic biases in the dataset:*** *Distribution* ***a-c.*** *of age groups (note referenced studies use different groupings for age),* ***d-f.*** *genders,* ***g.*** *ethnicities, and* ***h-j.*** *UK regions in study dataset in comparison to general population (Census 2021)[54], NHS Test and Trace users who took a PCR or LAMP test 25-02-2021 to 02-03-2022[41], participants in REACT-1 rounds 13-18[42–45]. Data for England only. Where study data demographic distributions are compared to NHS Test and Trace and REACT-1, only the study data subset from each respective recruitment channel are used. Groups are created according to how baseline data is reported. 'Unknown' categories are not displayed. Ethnicity data is not yet published for the 2021 UK Census at the time of writing. NHS Test and Trace ethnicity data by week is not publicly available. *Y&H = Yorkshire and The Humber.*

## 6. Ethics

This study has been approved by The National Statistician's Data Ethics Advisory Committee (reference NSDEC(21)01) and the Cambridge South NHS Research Ethics Committee (reference 21/EE/0036) and Nottingham NHS Research Ethics Committee (reference 21/EM/0067).

## 7. Code Availability

Code used to produce figures from the full dataset can be found here:
https://github.com/alan-turing-institute/Turing-RSS-Health-Data-Lab-Biomedical-Acoustic-Markers/

## 8. Acknowledgements

Authors gratefully acknowledge the contributions of staff from NHS Test and Trace Lighthouse Labs, REACT Study, Ipsos MORI, Studio24, Fujitsu Services Ltd. Authors in The Alan Turing Institute and Royal Statistical Society Health Data Lab gratefully acknowledge funding from Data, Analytics and Surveillance Group, a part of the UKHSA. This work was funded by The Department for Health and Social Care (Grant ref: 2020/045) with support from The Alan Turing Institute (EP/W037211/1) and in-kind support from The Royal Statistical Society. J.B. and R.A.M. acknowledge funding from the i-sense EPSRC IRC in Agile Early Warning Sensing Systems for Infectious Diseases and Antimicrobial Resistance EP/R00529X/1. A.T.C. acknowledges funding from the European Union's Horizon 2020 research and innovation programme under the Marie Skłodowska–Curie grant agreement No 801604. SP acknowledges funding from the Economic and Social Research Council (ESRC) [grant number ES/P000592/1].

## 9. Author contributions

J.B.: Conceptualization, Data curation, Formal Analysis, Investigation, Methodology, Project administration, Software, Visualization, Writing – original draft, Writing – review & editing

K.B.: Conceptualization, Data curation, Formal Analysis, Investigation, Methodology, Software, Validation, Visualization, Writing – original draft, Writing – review & editing

E.K.: Project administration, Writing – original draft, Writing – review & editing

H.C.: Conceptualization, Data curation, Formal Analysis, Investigation, Methodology, Software, Validation, Visualization, Writing – review & editing

S.P.: Funding acquisition, Investigation, Methodology, Project administration, Resources, Supervision, Writing – review & editing

A.T.C.: Investigation, Project administration, Resources, Supervision, Writing – review & editing

A.T.: Project administration, Resources, Software, Supervision, Validation, Visualization, Writing – original draft, Writing – review & editing

R.P.: Data curation, Investigation, Software, Supervision, Validation, Writing – review & editing

D.H.: Data curation, Investigation, Software, Supervision, Validation, Writing – original draft

S.E.: Funding acquisition, Project administration, Resources,, Writing – review & editing

L.B.: Investigation, Project administration, Resources, Writing – review & editing

J.M.: Supervision, Writing – review & editing

G.N.: Data curation, Formal Analysis, Investigation, Methodology, Validation, Writing – review & editing

I.K.: Conceptualization, Data curation, Formal Analysis, Investigation, Methodology, Software, Supervision, Writing – review & editing

V.K.: Conceptualization, Investigation, Methodology, Writing – review & editing

R.J.: Validation, Writing – review & editing

R.A.M.: Supervision, Writing – review & editing

P.D., S.R.: Supervision

B.S.: Conceptualization, Formal Analysis, Investigation, Methodology, Supervision, Writing – review & editing

S.G.: Formal Analysis, Funding acquisition, Methodology, Project administration, Supervision

D.P.: Formal Analysis, Supervision, Validation, Writing – review & editing

S.R.: Conceptualization, Formal Analysis, Investigation, Methodology, Project administration, Supervision, Writing – review & editing

J.P.: Conceptualization, Funding acquisition, Investigation, Project administration, Resources, Supervision, Writing – review & editing

T.T.: Conceptualization, Investigation, Resources, Supervision, Writing – review & editing

C.H.: Conceptualization, Investigation, Methodology, Project administration, Resources, Supervision, Writing – review & editing

## 10. Competing interests

None to declare.

| Prompt number | Prompt | Response type | Response options | Relevant columns in participant metadata | Relevant columns in audio metadata |
|---|---|---|---|---|---|
| 1 | Do you have any of the following symptoms? Select the options that currently apply to you. You can select multiple symptoms. | Able to select multiple options | Fever (a high temperature); Headache; Fatigue; Abdominal pain; Diarrhoea; Cough (any); A new continuous cough; A change to sense of smell or taste; Loss of taste; Shortness of breath; Runny or blocked nose (option available from 2021-08-11); Sore throat (option available from 2021-07-22); Other symptom(s) new to you in the last 2 weeks (option available from 2021-07-22); No symptoms; Prefer not to say | symptom_none, symptom_cough_any, symptom_new_continuous_cough, symptom_runny_or_blocked_nose, symptom_shortness_of_breath, symptom_sore_throat, symptom_abdominal_pain, symptom_diarrhoea, symptom_fatigue, symptom_fever_high_temperature, symptom_headache, symptom_change_to_sense_of_smell_or_taste, symptom_loss_of_taste, symptom_other, symptom_prefer_not_to_say | n/a |
| 2 | Do you smoke? | Able to select one option only | Never smoked; Ex-smoker; Current smoker (1 to 10 cigarettes per day); Current smoker (11 or more cigarettes per day); Current smoker (e-cigarettes or vapes only); Prefer not to say | smoker_status | n/a |
| 3 | Do you currently have any of the following conditions? Select the options that currently apply to you. You can select multiple options | Able to select multiple options | Asthma; COPD or Emphysema; Other respiratory condition; None of the above; Prefer not to say | respiratory_condition_none, respiratory_condition_asthma, respiratory_condition_copd_or_emphysema, respiratory_condition_other, respiratory_condition_prefer_not_to_say | n/a |
| 4 | What is your first language? If you have more than one first language, please choose the language that you use most often at home. | Able to select one option only | English; Welsh/Cymraeg; British sign language; Afrikaans; Akan; Albanian; Amharic; Arabic; Bengali (including Sylheti and Chatgaya); Bulgarian; Caribbean Creole (English-based); Chinese (Cantonese); Chinese (Mandarin); Czech; Danish; Dutch; Estonian; Finnish; French; German; Greek; Gujurati; Hebrew; Hindi; Hungarian; Igbo; Italian; Japanese; Korean; Krio; Kurdish; Latvian; Lingala; Lithuanian; Luganda; Malay; Malayalam; Maltese; Marathi; Nepalese; Pakistani Pahari (including Mirpuri and Potwari); Panjabi; Pashto; Persian/Farsi; Polish; Portuguese; Romani language (any); Romanian; Russian; Serbian/Croatian/Bosnian; Shona; Sinhala; Slovak; Slovenian; Somali; Spanish; Swahili/Kiswahili; Swedish; Tagalog/Filipino; Tamil; Telugu; Thai; Tigrinya; Turkish; Ukrainian; Urdu; Vietnamese; Yiddish; Yoruba; All other languages; Prefer not to say | language | n/a |
| 5 | What is your height? | *Able to select one option for units, and one category* | Feet and Inches; Centimetres; 1ft 7in to 1ft 11in; 1ft 11in to 2ft 3in; 2ft 3in to 2ft 7in; 2ft 7in to 2ft 11in; 2ft 11in to 3ft 2in; 3ft 3in to 3ft 6in; 3ft 7in to 3ft 10in; 3ft 11in to 4ft 2in; 4ft 3in to 4ft 6in; 4ft 7in to 4ft 10in; 4ft 11in to 5ft 2in; 5ft 2in to 5ft 6in; 5ft 6in to 5ft 10in; 5ft 10in to 6ft 2in; 6ft 2in to 6ft 6in; 6ft 6in to 6ft 10in; 6ft 10in to 7ft 2in; 7ft 2in to 7ft 6in; 7ft 6in to 7ft 10in; 50cm to 59cm; 60cm to 69cm; 70cm to 79cm; 80cm to 89cm; 90cm to 99cm; 100cm to 109cm; 110cm to 119cm; 120cm to 129cm; 130cm to 139cm; 140cm to 149cm; 150cm to 159cm; 160cm to 169cm; 170cm to 179cm; 180cm to 189cm; 190cm to 199cm; 200cm to 209cm; 210cm to 219cm; 220cm to 229cm; 230cm to 239cm; Prefer not to say | height | n/a |
| 6 | What is your weight? Choose your preferred unit of measurement, then select your weight. | *Able to select one option for units, and one category* | Stones and Pounds; Kilograms; 20kg to 29kg; 30kg to 39kg; 40kg to 49kg; 50kg to 59kg; 60kg to 69kg 70kg to 79kg; 80kg to 89kg; 90kg to 99kg; 100kg to 109kg; 110kg to 119kg 120kg to 129kg; 130kg to 139kg; 140kg to 149kg; 150kg to 159kg; 160kg to 169kg 170kg to 179kg; 180kg to 189kg; 190kg to 199kg; 200kg to 209kg; Prefer not to say | weight | n/a |
| 7 | Are you wearing a mask right now? | Able to select one option only | Yes; No; Prefer not to say | wearing_mask | n/a |
| 8 | Record a sentence “I love nothing more than an afternoon cream tea.” This sentence contains some key sounds (‘aaah’, ‘oooh’, ‘eee’) which may help to indicate your respiratory health Press the Record button, and read the following sentence: “I love nothing more than an afternoon cream tea.” Use the Stop button to stop recording. You will see an audio player, which you can use to playback your recording. (Press Record again, if you want to try again.) Once you are happy with your recording, press the Save recording and continue button. Read the following sentence: “I love nothing more than an afternoon cream tea.” Press the Record button to begin recording. | Audio recording | n/a | n/a | sentence_file_name, sentence_size, sentence_sample_rate, sentence_frames, sentence_channels, sentence_length, sentence_amplitude, sentence_snr, sentence_transcript, sentence_transcript_outlier_score |
| 9 | Record a 'ha' sound Please make this recording in a quiet environment. Press the Record button, and breathe out loudly three times, making a 'ha' sound, as if you were trying to fog up a window, or see your breath in cold weather Use the Stop button to stop recording. You will see an audio player, which you can use to playback your recording. (Press Record again, if you want to try again.) Once you are happy with your recording, press the Save recording and continue button. Breathe out loudly three times, making a 'ha' sound, as if you were trying to fog up a window, or see your breath in cold weather. Press the Record button to begin recording | Audio recording | n/a | n/a | exhalation_file_name, exhalation_size, exhalation_sample_rate, exhalation_frames, exhalation_channels, exhalation_length, exhalation_amplitude, exhalation_snr |
| 10 | Record a cough. Coughing is a potential risk to others around you. Make sure you are alone in a room or vehicle when coughing. For this recording, move an arm's length away from your desktop computer, laptop, phone or tablet. Press the Record button, and cough, forcing a cough if it doesn't come naturally. Use the Stop button to stop recording. You will see an audio player, which you can use to playback your recording. (Press Record again, if you want to try again.) Once you are happy with your recording, press the Save recording and continue button. Cough once — with your desktop computer, laptop, phone or tablet an arm's length away from you | Audio recording | n/a | n/a | cough_file_name, cough_size, cough_sample_rate, cough_frames, cough_channels, cough_length, cough_amplitude, cough_snr |
| 11 | Record three coughs Coughing is a potential risk to others around you. Make sure you are alone in a room or vehicle when coughing. For this recording, move an arm's length away from your desktop computer, laptop, phone or tablet. Press the Record button, and cough three times. Use the Stop button to stop recording. (Press Record again, if you want to try again.)You will see an audio player, which you can use to playback your recording. Once you are happy with your recording, press the Save recording and continue button. Cough three times — with your desktop computer, laptop, phone or tablet an arm's length away from you. | Audio recording | n/a | n/a | three_cough_file_name, three_cough_size, three_cough_sample_rate, three_cough_frames, three_cough_channels, three_cough_length, three_cough_amplitude, three_cough_snr |

Supplementary Table 2: Granular Recruitment Information

| Ref | Recruitment approach | Case type | Start date | End date | Engagement method | Total individuals reached | Description |
|---|---|---|---|---|---|---|---|
| 1 | NHS Test and Trace | Positive and Negative | Feb-21 | Mar-21 | Leaflet | ~18,000 | Pilot phase |
| 2 | NHS Test and Trace | Positive | Mar-21 | Mar-22 | SMS | 474,097 | After a positive test is processed at a Test and Trace lab, or rapid antigen (lateral flow) result is reported, a sample of participants who have not already been contacted by NHS Test and Trace in support of another study are sent a text message to invite them to take part. |
| 3 | NHS Test and Trace | Positive | Mar-21 | Mar-22 | Phone call | 16,477 | A sample of individuals engaged through 'Ref 1' above receive a follow-up phone call to invite them to participate in the study, prioritising individuals who undertook PCR tests over lateral flow. |
| 4 | NHS Test and Trace | Negative | Sep-21 | Oct-21 | Email | 43,161 | After a negative test is processed at a Test and Trace lab, or rapid antigen (lateral flow) result is reported, a sample of participants who have not already been contacted by NHS Test and Trace in support of another study are sent an email to invite them to take part. |
| 5 | NHS Test and Trace | Negative | Nov-21 | Mar-22 | SMS | 162,489 | After a negative test is processed at a Test and Trace lab, or rapid antigen (lateral flow) result is reported, a sample of participants who have not already been contacted by NHS Test and Trace in support of another study are sent a text message to invite them to take part. |
| 6 | NHS Test and Trace | Positive and Negative | Nov-21 | Mar-22 | 'Voice of the Customer' web survey | Unable to define | After conducting a COVID-19 test, a sample of individuals are invited to provide feedback to NHS Test and Trace. At the end of the feedback survey, they are invited to take part in this study and are provided a link to the participant information sheet. |
| 7 | REACT Round 13 | Positive and Negative | 27-06-21 | 14-07-21 | Email | Unable to define | All participants in the REACT study are invited to opt-in to be informed about other COVID-19 studies they can support. Those who do opt-in are then invited to take part in this study after they have completed the REACT swab and survey |
| 8 | REACT Round 14 | Positive and Negative | 12-09-21 | 29-09-21 | Email | 49,415 | All participants in the REACT study are invited to opt-in to be informed about other COVID-19 studies they can support. Those who do opt-in are then invited to take part in this study after they have completed the REACT swab and survey |
| 9 | REACT Round 15 | Positive and Negative | 21-10-21 | 05-11-21 | Email | 62,467 | All participants in the REACT study are invited to opt-in to be informed about other COVID-19 studies they can support. Those who do opt-in are then invited to take part in this study after they have completed the REACT swab and survey |
| 10 | REACT Round 16 | Positive and Negative | 26-11-21 | 16-12-21 | Email | 59,184 | All participants in the REACT study are invited to opt-in to be informed about other COVID-19 studies they can support. Those who do opt-in are then invited to take part in this study after they have completed the REACT swab and survey |
| 11 | REACT Round 17 | Positive and Negative | 08-01-22 | 22-01-22 | Email | 66,387 | All participants in the REACT study are invited to opt-in to be informed about other COVID-19 studies they can support. Those who do opt-in are then invited to take part in this study after they have completed the REACT swab and survey |
| 12 | REACT Round 18 | Positive and Negative | 11-02-22 | 03-03-22 | Email | 58,040 | All participants in the REACT study are invited to opt-in to be informed about other COVID-19 studies they can support. Those who do opt-in are then invited to take part in this study after they have completed the REACT swab and survey |

Supplementary Table 3: Participant Metadata Data Dictionary

| Column Name | Data Type | Description | Source | Completeness including "Prefer not to say", "Unknown", "Uknown/Void" (%) | Completeness excluding "Prefer not to say", "Unknown", "Uknown/Void" (%) | Range of Values | Notes |
|---|---|---|---|---|---|---|---|
| participant_identifier | object | 7-character identifier of the format p012345, denoting an individual participant. Can be used to link metadata variables across tables and to audio files. | Post-processing | 100.00 | 100.00 | p00000 - p73000 | In no particular order |
| survey_phase | object | Alpha (2021-03-01 to 2021-08-12) or beta ( 2021-08-13 to 2022-03-07) phase where different data gathering solution was used | Post-processing | 100.00 | 100.00 | "alpha", "beta" | |
| recruitment_source | object | Indicated whether participant was recruited via NHS Test and Trace or REACT survey round, and indicates source of associated metadata | Post-processing | 100.00 | 100.00 | "Test and Trace", "REACT Study Round 13", "REACT Study Round 14", "REACT Study Round 15", "REACT Study Round 16", "REACT Study Round 17", "REACT Study Round 18" | |
| submission_date | date | Date that participant contributed to the study digital survey | Study Survey | 100.00 | 100.00 | 2021-03-01 - 2022-03-07 | |
| submission_hour | int | Hour of day that the particpant contributed to the study digital survey | Study Survey | 100.00 | 100.00 | 0 - 23 | Alpha survey: final audio recording submission time Beta survey: survey start time |
| submission_delay | float | Estimated delay rounded to days between the participant taking the PCR (based on swab date/time estimated by participant or test site records, with missing values estimated using date PCR processed in lab minus 48 hours) and the time that the participant made the submission to the study survey, including audio recording | Post-processing | 99.99 | 99.99 | -27.0 - 539.0 | Where the "covid_test_date" for the test is not available, we use "covid_test_processed_date" minus 48 hours to estimate the submission delay |
| age | object (mixed int and string) | Participant age in years at time of taking COVID-19 test | NPEx; REACT | 100.00 | 100.00 | 18 - "94+" | Calculated from year of birth for REACT |
| gender | object | Participant gender | NPEx; REACT | 100.00 | 99.96 | "Female", "Male", "Unknown" | Categories for REACT and Test and Trace limited to Female/Male |
| region_name | object | Name of UK region where participant was recorded as residing at the time of the study. Derived from pseudonymised_local_authority_code before pseudonymisation | Post-processing | 99.97 | 99.97 | "East Midlands", "East of England", "London", "North East", "North West", "Scotland", "South East", "South West", "Wales", "West Midlands", "Yorkshire and The Humber" | |
| region_code | object | Office for National Statistics reference code of the UK region where participant was recorded as residing at the time of the study. Derived from pseudonymised_local_authority_code before pseudonymisation. Can be used to map statistics to region boundaries. | Post-processing | 99.97 | 99.97 | "E12000001", "E12000002", "E12000003", "E12000004", "E12000005", "E12000006", "E12000007", "E12000008", "E12000009", "S92000003", "W92000004" | |
| pseudonymised_local_authority | object | Pseudonymised code of the local authority in which the participant was recorded as residing at the time of the study. As declared through test provider. Re-formatted to 2021 LAD boundaries before pseudonymisation. | NPEx; REACT; Post-processing | 99.97 | 99.97 | "LAD00XXX" | |
| ethnicity | object | Participant ethnicity | NPEx; REACT | 100.00 | 100.00 | "African", "Another Asian background", "Another Black background", "Another Mixed background", "Another White background", "Another ethnic background", "Arab", "Asian and White", "Asian or Asian British", "Bangladeshi", "Black African and White", "Black Caribbean and White", "Caribbean", "Chinese", "Indian", "Irish", "Irish Traveller or Gypsy", "Jewish", "No response", "Pakistani", "White British" | REACT value names re-mapped to match NPEX where possible |

## Supplementary Table 3: Participant Metadata Data Dictionary

| | | | | | | | |
|---|---|---|---|---|---|---|---|
| language | object | Participant first language (or language most often spoken at home of more than one first language) | Study survey (prompt 4) | 100.00 | 99.97 | "Afrikaans", "Akan", "Albanian", "All other languages", "Arabic", "Bengali (including Sylheti and Chatgaya)", "Bulgarian", "Chinese (Cantonese)", "Chinese (Mandarin)", "Czech", "Danish", "Dutch", "English", "Finnish", "French", "German", "Greek", "Gujrati", "Hebrew", "Hindi", "Hungarian", "Igbo", "Italian", "Japanese", "Korean", "Latvian", "Lithuanian", "Low count language N", "Low count language O", "Low count language P", "Low count language Q", "Low count language R", "Low count language S", "Low count language T", "Low count language U", "Low count language V", "Low count language W", "Low count language X", "Low count language Y", "Low count language Z", "Malay", "Malayalam", "Marathi", "Nepalese", "Pahari (including Mirpuri and Potwari)", "Pakistani", "Panjabi", "Pashto", "Persian/Farsi", "Polish", "Portuguese", "Prefer not to say", "Romanian", "Russian", "Serbian/Croatian/Bosnian", "Shona", "Sinhala", "Slovak", "Somali", "Spanish", "Swedish", "Tagalog/Filipino", "Tamil", "Telugu", "Thai", "Turkish", "Ukrainian", "Urdu", "Welsh/Cymraeg", "Yoruba" | Language options from ONS 2011 UK Census |
| height | object (mixed int and string) | Participant height in centimetres chosen from categories. Some categories availble in imperial units were converted. All categories rounded to nearest 10 centimetres to category mid-point | Study survey (prompt 5) | 99.97 | 99.40 | "<=90" - 230, "Prefer not to say" | Nearest 10 cm |
| weight | float | Participant height in kilograms chosen from categories. Some categories availble in imperial units were converted. All categories rounded to nearest 10 kilograms to category mid-point | Study survey (prompt 6) | 99.94 | 97.38 | 20 - 200, "Prefer not to say" | Nearest 10 kg |
| wearing_mask | object | Whether the participant is wearing a mask at the time of completing the study survey and recording audio | Study survey (prompt 7) | 100.00 | 99.96 | "No", "Yes", "Prefer not to say" | |
| covid_test_result | object | Result of the participant's SARS-CoV-2 test | NPEx; REACT | 100.00 | 99.76 | "Positive", "Negative", "Unknown/Void" | REACT value names re-mapped to match NPEX |
| covid_test_method | object | Method of COVID-19 test | NPEx; REACT | 99.96 | 99.95 | "LAMP", "LFT", "PCR", "RT-qPCR", "Unknown", "ePCR" | |
| covid_test_date | date | The date the participant took the test, sometimes approximated as declared by the individual | NPEx; REACT; Post-processing | 95.01 | 95.01 | 2020-08-27 - 2022-03-07 | Calculated from several REACT-provided categories |
| covid_test_processed_date | date | The date the swab was processed (used to estimate "submission_delay" when "covid_test_date" not available) | NPEx; REACT | 99.93 | 99.93 | 2020-08-29 - 2022-03-07 | |
| covid_test_lab_code | bool | Code indicating whether the laboratory (that processed the PCR or LAMP COVID-19 test) reported test result issues during Sept/Oct 2021. True for laboratories with issues. False for all others. | NPEx; REACT | 100.00 | 100.00 | True, False | "covid_test_result" not deemed reliable if True |
| covid_ct_value | float | Returns the lowest Ct value found across all Ct values available for a given test result (multiple values correspond to different target genes) | NPEx; REACT | 25.20 | 25.20 | 5.0 - 48.0 | |
| covid_ct_gene | object | Indicates the gene target for the ‘covid_ct_value’ identified | NPEx; REACT | 25.20 | 25.20 | "E gene", "N gene", "ORF1ab gene", "S gene" | |
| covid_ct_mean | object | Returns the mean Ct value found across all Ct values available for a given test result (multiple values correspond to different target genes) | NPEx; REACT | 14.62 | 14.62 | 8.28 - 36.44 | Genes in scope for this calculation: ORF1ab gene, S gene, N gene (where available) |
| covid_viral_load | float | Estimated viral load for a given COVID-19 PCR test result, approximated from covid_ct_mean. | NPEx; REACT | 14.62 | 14.62 | 1.11 - 1921286462.35 | |
| covid_viral_load_category | object | Range category of covid_viral_load (low, medium, high) where >1,000,000 viral RNA copies/ml is an approximate threshold for transmissibility | NPEx; REACT | 14.62 | 14.62 | High viral load (>1M copies/ml)", "Low viral load (10K-1M copies/ml)", "Minimal viral load (< 10K copies/ml)" | Formula used: 10.0 ** (12.0-0.328 * covid_ct_mean). This is a very rough estimate of viral load, and unlikely to be reliable for individual results |
| covid_vaccine_doses | object | Number of COVID-19 vaccine doses the participant has received | NPEx; REACT | 94.55 | 94.55 | 0, 1, 2, ">2" | |
| covid_vaccine_period | object | Number of days since last COVID-19 vaccine | NPEx; REACT | 91.61 | 91.61 | "Less than 7 days", "7 to 14 days", "More than 14 days" | |
| symptom_onset | float | Estimated number of days between onset of symptoms and completion of the study survey, including audio recording. Estimated by the participant. | NPEx; REACT | 36.05 | 36.05 | -1.0 - 2513.0 | REACT: 1 day (24 hours) added to 'SYMPTST' to account for delay in completing the study survey after completion of REACT survey where this value is declared NPEX: Calculated by subtracting 'DateOfOnset' from the survey 'submission_time' |
| symptom_none | int | Binary value indicating no symptoms (1) or any symptoms (0) as indicated by particpant | Study survey (prompt 1) | 100.00 | 100.00 | 0, 1 | |
| symptom_cough_any | int | Binary value (1 or 0) indicating prescence or absence of symptom (any cough) as indicated by participant | Study survey (prompt 1) | 100.00 | 100.00 | 0, 1 | |
| symptom_new_continuous_cough | int | Binary value (1 or 0) indicating prescence or absence of symptom (a new continuous cough) as indicated by participant | Study survey (prompt 1) | 100.00 | 100.00 | 0, 1 | |
| symptom_runny_or_blocked_nose | int | Binary value (1 or 0) indicating prescence or absence of symptom (runny or blocked nose) as indicated by participant | Study survey (prompt 1) | 100.00 | 100.00 | 0, 1 | Option not available to all participants as was added 11/08/21 |
| symptom_shortness_of_breath | int | Binary value (1 or 0) indicating prescence or absence of symptom (shortness of breath) as indicated by participant | Study survey (prompt 1) | 100.00 | 100.00 | 0, 1 | |
| symptom_sore_throat | int | Binary value (1 or 0) indicating prescence or absence of symptom (sore throat) as indicated by participant | Study survey (prompt 1) | 100.00 | 100.00 | 0, 1 | Option not available to all participants as was added 22/07/21 |
| symptom_abdominal_pain | int | Binary value (1 or 0) indicating prescence or absence of symptom (abdominal pain) as indicated by participant | Study survey (prompt 1) | 100.00 | 100.00 | 0, 1 | |
| symptom_diarrhoea | int | Binary value (1 or 0) indicating prescence or absence of symptom (diarrhoea) as indicated by participant | Study survey (prompt 1) | 100.00 | 100.00 | 0, 1 | |
| symptom_fatigue | int | Binary value (1 or 0) indicating prescence or absence of symptom (fatigue) as indicated by participant | Study survey (prompt 1) | 100.00 | 100.00 | 0, 1 | |
| symptom_fever_high_temperature | int | Binary value (1 or 0) indicating prescence or absence of symptom (fever or a high temperature) as indicated by participant | Study survey (prompt 1) | 100.00 | 100.00 | 0, 1 | |
| symptom_headache | int | Binary value (1 or 0) indicating prescence or absence of symptom (headache) as indicated by participant | Study survey (prompt 1) | 100.00 | 100.00 | 0, 1 | |
| symptom_change_to_sense_of_smell_or_taste | int | Binary value (1 or 0) indicating prescence or absence of symptom (a change to sense of smell or taste) as indicated by participant | Study survey (prompt 1) | 100.00 | 100.00 | 0, 1 | |
| symptom_loss_of_taste | int | Binary value (1 or 0) indicating prescence or absence of symptom (loss of sense of taste) as indicated by participant | Study survey (prompt 1) | 100.00 | 100.00 | 0, 1 | |
| symptom_other | int | Binary value (1 or 0) indicating prescence or absence of symptom (other symptom not included in available options) as indicated by participant | Study survey (prompt 1) | 100.00 | 100.00 | 0, 1 | Option not available to all participants as added 22/07/21 |
| symptom_prefer_not_to_say | int | Binary value (1 or 0) indicating a preference of the particpant not to disclose symptoms | Study survey (prompt 1) | 100.00 | 100.00 | 0, 1 | |
| smoker_status | object | String indiciating whether the participant smokes or used to smoke, and if so to what extent | Study survey (prompt 2) | 100.00 | 99.31 | "Current smoker (1 to 10 cigarettes per day)", "Current smoker (11 or more cigarettes per day)", "Current smoker (e-cigarettes or vapes only)", "Ex-smoker", "Never smoked", "Prefer not to say" | |
| respiratory_condition_none | int | Binary value (1 or 0) indicating prescence or absence of condition as indicated by participant | Study survey (prompt 3) | 100.00 | 100.00 | 0, 1 | |
| respiratory_condition_asthma | int | Binary value (1 or 0) indicating prescence or absence of condition (asthma) as indicated by participant | Study survey (prompt 3) | 100.00 | 100.00 | 0, 1 | |
| respiratory_condition_copd_or_emphysema | int | Binary value (1 or 0) indicating prescence or absence of condition (COPD or emphysema) as indicated by participant | Study survey (prompt 3) | 100.00 | 100.00 | 0, 1 | |
| respiratory_condition_other | int | Binary value (1 or 0) indicating prescence or absence of condition (any other respiratory health condition not included in available options) as indicated by participant | Study survey (prompt 3) | 100.00 | 100.00 | 0, 1 | |
| respiratory_condition_prefer_not_to_say | int | Binary value (1 or 0) indicating a preference of the particpant not to disclose respiratory condition status | Study survey (prompt 3) | 100.00 | 100.00 | 0, 1 | |
| influenza_a_test_result | object | PCR test results for Influenza A (REACT rounds 16-18 only) | REACT | 30.86 | 27.20 | "Negative", "Positive", "Unknown/Void" | |
| influenza_a_ct_value | float | Cycle threshold value for the associated Influenza A PCR test result if positive | REACT | 27.23 | 27.23 | 0.00 - 40.00 | |
| influenza_b_test_result | object | PCR test results for Influenza B (REACT rounds 16-18 only) | REACT | 30.86 | 27.20 | "Negative", "Positive", "Unknown/Void" | |
| influenza_b_ct_value | float | Cycle threshold value for the associated Influenza B PCR test result if positive | REACT | 27.23 | 27.23 | 0.00 - 40.00 | |
| influenza_vaccine | object | Influenza vaccination status | REACT | 41.50 | 41.50 | "Yes", "No" | Vaccination status since 2021-09-01 |
| influenza_vaccine_date | date | Date of influenza vaccine if vaccinated | REACT | 28.19 | 28.19 | 2021-01-05 - 2022-02-17 | |

| Column Name | Data Type | Description | Source | Completeness (%) | Range of Values (to 2 decimal places) |
|---|---|---|---|---|---|
| participant_identifier | Object | 7-character identifier of the format p012345, denoting an individual participant. Can be used to link metadata variables across tables and to audio files. | Post-processing | 100.00 | p00000 - p73000 |
| exhalation_file_name | Object | File name for the audio file associated with three successive exhalation sounds from the participant | Study Survey (prompt 9); Post-processing | 100.00 | Unique file name |
| exhalation_size | Float | Size in bytes of the audio file associated with three successive exhalation sounds from the participant | Post-processing | 99.25 | 2.0 -  3383340.0 |
| exhalation_sample_rate | Float | The sample rate in kilohertz of the audio file associated with three successive exhalation sounds from the participant | Post-processing | 98.77 | 8000.0 - 192000.0 |
| exhalation_frames | Float | The number of frames of the audio file associated with three successive exhalation sounds from the participant | Post-processing | 98.77 | 4096.0 - 1691648.0 |
| exhalation_channels | Float | The number of channels of the audio file associated with three successive exhalation sounds from the participant | Post-processing | 98.77 | 1.0 - 1.0 |
| exhalation_length | Float | Length in seconds of the audio file associated with three successive exhalation sounds from the participant | Post-processing | 98.77 | 0.09 - 30.21 |
| exhalation_amplitude | Float | Difference between the maximum and minumum of the absolute audio signal of the audio file associated with three successive exhalation sounds from the participant | Post-processing | 98.77 | 0.0 - 32768.0 |
| exhalation_snr | Float | Ratio between the mean and the standard deviation of the absolute audio signal of the audio file associated with three successive exhalation sounds from the participant | Post-processing | 98.54 | 0.00 - 1.81 |
| sentence_file_name | Object | File name for the audio file associated with a read sentence from the participant | Study Survey (prompt 8); Post-processing | 100.00 | Unique file name |
| sentence_size | Float | Size in bytes of the audio file associated with a read sentence from the participant | Post-processing | 99.63 | 2.0 - 3293228.0 |
| sentence_sample_rate | Float | The sample rate in kilohertz of the audio file associated with a read sentence from the participant | Post-processing | 99.36 | 8000.0 - 192000.0 |
| sentence_frames | Float | The number of frames of the audio file associated with a read sentence from the participant | Post-processing | 99.36 | 4096.0 - 1646592.0 |
| sentence_channels | Float | The number of channels of the audio file associated with a read sentence from the participant | Post-processing | 99.36 | 1.0 - 1.0 |
| sentence_length | Float | Length in seconds of the audio file associated with a read sentence from the participant | Post-processing | 99.36 | 0.09 - 72.19 |
| sentence_amplitude | Float | Difference between the maximum and minumum of the absolute audio signal of the audio file associated with a read sentence from the participant | Post-processing | 99.36 | 0.0 - 32768.0 |
| sentence_snr | Float | Ratio between the mean and the standard deviation of the absolute audio signal of the absolute audio signal of the audio file associated with a read sentence from the participant | Post-processing | 99.08 | 0.00 - 2.41 |
| sentence_transcript | Object | Transcript of the audio file associated with a read sentence from the participant, as inferred by the Speech to Text Transformer, S2T model | Post-processing | 99.36 | 11929 unique transcripts |
| sentence_transcript_outlier_score | Float | Outlier score of the sentence_transcript, where the transcript has been converted to an embedding using a test-to-embedding transformer, MPNet Transformer, and embedding similarities quantified using a Support Vector Machine. Higher values are more similar to the mean. | Post-processing | 99.08 | 36400.28 - 36475.12 |
| cough_file_name | Object | File name for the audio file associated with a single volitional cough from the participant | Study Survey (prompt 10); Post-processing | 100.00 | Unique file name |
| cough_size | Float | Size in bytes of the audio file associated with a single volitional cough from the participant | Post-processing | 99.94 | 2.0 - 3219500.0 |
| cough_sample_rate | Float | The sample rate in kilohertz of the audio file associated with a single volitional cough from the participant | Post-processing | 99.64 | 8000.0 - 192000.0 |
| cough_frames | Float | The number of frames of the audio file associated with a single volitional cough from the participant | Post-processing | 99.64 | 4096.0 - 1609728.0 |
| cough_channels | Float | The number of channels of the audio file associated with a single volitional cough from the participant | Post-processing | 99.64 | 1.0 - 1.0 |
| cough_length | Float | Length in seconds of the audio file associated with a single volitional cough from the participant | Post-processing | 99.64 | 0.09 - 48.90 |
| cough_amplitude | Float | Difference between the maximum and minumum of the absolute audio signal of the audio file associated with a single volitional cough from the participant | Post-processing | 99.64 | 0.0 - 32768.0 |
| cough_snr | Float | Ratio between the mean and the standard deviation of the absolute audio signal of the audio file associated with a single volitional cough from the participant | Post-processing | 99.40 | 0.00 - 1.81 |
| three_cough_file_name | Object | File name for the audio file associated with a three successive volitional coughs from the participant | Study Survey (prompt 11); Post-processing | 100.00 | Unique file name |
| three_cough_size | Float | Size in bytes of the audio file associated with three successive volitional coughs from the participant | Post-processing | 99.58 | 2.0 - 3981356.0 |
| three_cough_sample_rate | Float | The sample rate in kilohertz of the audio file associated with three successive volitional coughs from the participant | Post-processing | 99.08 | 8000.0 - 192000.0 |
| three_cough_frames | Float | The number of frames of the audio file associated with three succressive volitional coughs from the participant | Post-processing | 99.08 | 4096.0 - 1990656.0 |
| three_cough_channels | Float | The number of channels of the audio file associated with three successive volitional coughs from the participant | Post-processing | 99.08 | 1.0 - 1.0 |
| three_cough_length | Float | Length in seconds of the audio file associated with three successive volitional coughs from the participant | Post-processing | 99.08 | 0.09 - 64.0 |
| three_cough_amplitude | Float | Difference between the maximum and minimum of the absolute audio signal of the audio file associated with three successive volitional coughs from the participant | Post-processing | 99.08 | 0.0 - 32768.0 |
| three_cough_snr | Float | Ratio between the mean and the standard deviation of the absolute audio signal of the audio file associated with three successive volitional coughs from the participant | Post-processing | 98.86 | 0.00 - 1.81 |
| missing_audio | Boolean | ‘True’ if one or more audio files is 44 bytes or smaller, or file is not present | Post-processing | 100.00 | True, False |

| Column Name | Data Type | Description | Source | Completeness including "Prefer not to say", "Unknown", "Uknown/Void" (%) | Completeness excluding "Prefer not to say", "Unknown", "Uknown/Void" (%) | Range of Values | Notes |
|---|---|---|---|---|---|---|---|
| participant_identifier | object | 8-character identifier of the format P0123456, denoting an individual participant. Can be used to link metadata variables across tables and to audio files. | Post-processing | 100.00 | 100.00 | PXXXXXXX | In no particular order |
| survey_phase | object | Alpha (2021-03-01 to 2021-08-12) or beta ( 2021-08-13 to 2022-03-07) phase where different data gathering solution was used | Post-processing | 100.00 | 100.00 | "alpha", "beta" | |
| recruitment_source | object | Indicated whether participant was recruited via NHS Test and Trace or REACT survey round, and indicates source of associated metadata. | Post-processing | 100.00 | 100.00 | "Test and Trace", "REACT" | |
| submission_date | int | Date that participant contributed to the study digital survey, indexed to a random date with +-10 days random noise. All dates included in the metadata are indexed to the same random date for comparison. | Study Survey | 100.00 | 100.00 | 218 - 598 | |
| submission_delay | float | Estimated delay rounded to days between the participant taking the PCR (based on swab date/time estimated by participant or test site records, with missing values estimated using date PCR processed in lab minus 48 hours) and the time that the participant made the submission to the study survey, including audio recording. | Post-processing | 99.99 | 99.99 | -27.0 - 539.0 | Where the "covid_test_date" for the test is not available, we use "covid_test_processed_date" minus 48 hours to estimate the submission delay |
| age | object | Participant age in years at time of taking COVID-19 test. | NPEx; REACT | 100.00 | 100.00 | "18-44", "45-64", "65+" | Calculated from year of birth for REACT |
| gender | object | Participant gender. | NPEx; REACT | 99.96 | 99.96 | "Female", "Male" | Categories for REACT and Test and Trace limited to Female/Male |
| region_name | object | Pseudonymised name of UK region where participant was recorded as residing at the time of the study. | Post-processing | 99.94 | 99.94 | "B", "E", "K", "M", "N", "P", "S", "T", "W" | |
| wearing_mask | object | Whether the participant is wearing a mask at the time of completing the study survey and recording audio. | Study survey (prompt 7) | 100.00 | 99.96 | "No", "Yes", "Prefer not to say" | |
| covid_test_result | object | Result of the participant's SARS-CoV-2 test. Value is None if the result was invalid. | NPEx; REACT | 99.65 | 99.65 | "Positive", "Negative" | REACT value names re-mapped to match NPEX |
| covid_test_method | object | Method of COVID-19 test. | NPEx; REACT | 99.96 | 99.95 | "LAMP", "LFT", "PCR", "RT-qPCR", "Unknown", "ePCR" | |
| covid_test_date | date | The date the participant took the test, sometimes approximated as declared by the individual, indexed to a random date with +-10 days random noise. . All dates included in the metadata are indexed to the same random date for comparison. | NPEx; REACT; Post-processing | 95.01 | 95.01 | 24 - 594 | Calculated from several REACT-provided categories |
| covid_test_processed_date | date | The date the swab was processed (used to estimate "submission_delay" when "covid_test_date" not available), indexed to a random date with +-10 days random noise. All dates included in the metadata are indexed to the same random date for comparison. | NPEx; REACT | 99.93 | 99.93 | 26 - 595 | |
| covid_ct_value | float | Returns the lowest Ct value found across all Ct values available for a given test result (multiple values correspond to different target genes) | NPEx; REACT | 25.20 | 25.20 | 5.0 - 48.0 | |
| covid_ct_gene | object | Indicates the gene target for the ‘covid_ct_value’ identified | NPEx; REACT | 25.20 | 25.20 | "E gene", "N gene", "ORF1ab gene", "S gene" | |
| covid_ct_mean | float | Returns the mean Ct value found across all Ct values available for a given test result (multiple values correspond to different target genes) | NPEx; REACT | 14.62 | 14.62 | 8.28 - 36.44 | Genes in scope for this calculation: ORF1ab gene, S gene, N gene (where available) |
| covid_viral_load | float | Estimated viral load for a given COVID-19 PCR test result, approximated from covid_ct_mean. | NPEx; REACT | 14.62 | 14.62 | 1.11 - 1921286462.35 | |
| covid_viral_load_category | object | Range category of covid_viral_load (low, medium, high) where >1,000,000 viral RNA copies/ml is an approximate threshold for transmissibility | NPEx; REACT | 14.62 | 14.62 | High viral load (>1M copies/ml)", "Low viral load (10K-1M copies/ml)", "Minimal viral load (< 10K copies/ml)" | Formula used: 10.0 ** (12.0-0.328 * covid_ct_mean). This is a very rough estimate of viral load, and unlikely to be reliable for individual results |
| symptom_onset | float | Estimated number of days between onset of symptoms and completion of the study survey, including audio recording. Estimated by the participant. | NPEx; REACT | 36.05 | 36.05 | -1.0 - 2513.0 | REACT: 1 day (24 hours) added to 'SYMPTST' to account for delay in completing the study survey after completion of REACT survey where this value is declared NPEX: Calculated by subtracting 'DateOfOnset' from the survey 'submission_time' |
| symptom_none | int | Binary value indicating no symptoms (1) or any symptoms (0) as indicated by particpant | Study survey (prompt 1) | 100.00 | 100.00 | 0, 1 | |
| symptom_cough_any | int | Binary value (1 or 0) indicating prescence or absence of symptom (any cough) as indicated by participant | Study survey (prompt 1) | 100.00 | 100.00 | 0, 1 | |
| symptom_new_continuous_cough | int | Binary value (1 or 0) indicating prescence or absence of symptom (a new continuous cough) as indicated by participant | Study survey (prompt 1) | 100.00 | 100.00 | 0, 1 | |
| symptom_runny_or_blocked_nose | int | Binary value (1 or 0) indicating prescence or absence of symptom (runny or blocked nose) as indicated by participant | Study survey (prompt 1) | 100.00 | 100.00 | 0, 1 | Option not available to all participants as was added 11/08/21 |
| symptom_shortness_of_breath | int | Binary value (1 or 0) indicating prescence or absence of symptom (shortness of breath) as indicated by participant | Study survey (prompt 1) | 100.00 | 100.00 | 0, 1 | |
| symptom_sore_throat | int | Binary value (1 or 0) indicating prescence or absence of symptom (sore throat) as indicated by participant | Study survey (prompt 1) | 100.00 | 100.00 | 0, 1 | Option not available to all participants as was added 22/07/21 |
| symptom_abdominal_pain | int | Binary value (1 or 0) indicating prescence or absence of symptom (abdominal pain) as indicated by participant | Study survey (prompt 1) | 100.00 | 100.00 | 0, 1 | |
| symptom_diarrhoea | int | Binary value (1 or 0) indicating prescence or absence of symptom (diarrhoea) as indicated by participant | Study survey (prompt 1) | 100.00 | 100.00 | 0, 1 | |
| symptom_fatigue | int | Binary value (1 or 0) indicating prescence or absence of symptom (fatigue) as indicated by participant | Study survey (prompt 1) | 100.00 | 100.00 | 0, 1 | |
| symptom_fever_high_temperature | int | Binary value (1 or 0) indicating prescence or absence of symptom (fever or a high temperature) as indicated by participant | Study survey (prompt 1) | 100.00 | 100.00 | 0, 1 | |
| symptom_headache | int | Binary value (1 or 0) indicating prescence or absence of symptom (headache) as indicated by participant | Study survey (prompt 1) | 100.00 | 100.00 | 0, 1 | |
| symptom_change_to_sense_of_smell_or_taste | int | Binary value (1 or 0) indicating prescence or absence of symptom (a change to sense of smell or taste) as indicated by participant | Study survey (prompt 1) | 100.00 | 100.00 | 0, 1 | |
| symptom_loss_of_taste | int | Binary value (1 or 0) indicating prescence or absence of symptom (loss of sense of taste) as indicated by participant | Study survey (prompt 1) | 100.00 | 100.00 | 0, 1 | |
| symptom_other | int | Binary value (1 or 0) indicating prescence or absence of symptom (other symptom not included in available options) as indicated by participant | Study survey (prompt 1) | 100.00 | 100.00 | 0, 1 | Option not available to all participants as added 22/07/21 |
| symptom_prefer_not_to_say | int | Binary value (1 or 0) indicating a preference of the particpant not to disclose symptoms | Study survey (prompt 1) | 100.00 | 100.00 | 0, 1 | |
| smoker_status | object | String indiciating whether the participant smokes or used to smoke, and if so to what extent | Study survey (prompt 2) | 100.00 | 99.31 | "Current smoker (1 to 10 cigarettes per day)", "Current smoker (11 or more cigarettes per day)", "Current smoker (e-cigarettes or vapes only)", "Ex-smoker", "Never smoked", "Prefer not to say" | |
| respiratory_condition_asthma | float | Binary value (1 or 0) indicating prescence or absence of condition (asthma) as indicated by participant. Is np.nan if no response. | Study survey (prompt 3) | 99.75 | 99.75 | 0, 1 | |
| respiratory_condition_other | float | Binary value (1 or 0) indicating prescence or absence of condition (any other respiratory health condition not included in available options) as indicated by participant. Is np.nan if no response. | Study survey (prompt 3) | 99.75 | 99.75 | 0, 1 | |
| influenza_a_test_result | object | PCR test results for Influenza A (REACT rounds 16-18 only) | REACT | 30.86 | 27.20 | "Negative", "Positive", "Unknown/Void" | |
| influenza_a_ct_value | float | Cycle threshold value for the associated Influenza A PCR test result if positive | REACT | 27.23 | 27.23 | 0.00 -  40.00 | |
| influenza_b_test_result | object | PCR test results for Influenza B (REACT rounds 16-18 only) | REACT | 30.86 | 27.20 | "Negative", "Positive", "Unknown/Void" | |
| influenza_b_ct_value | float | Cycle threshold value for the associated Influenza B PCR test result if positive | REACT | 27.23 | 27.23 | 0.00 - 40.00 | |